\documentclass[biblatex]{revtex4-2}
\usepackage[utf8]{inputenc}
\usepackage[english]{babel}
\usepackage{natbib}
\usepackage{amsmath}
\usepackage{graphicx}
\usepackage{float}
\usepackage{physics}
\begin{document}

%TC:ignore

\title{Towards a statistical physics of dating apps}
\author{Fabrizio Olmeda} 
\email{fabrizio@pks.mpg.de}
\affiliation{Max Planck Institute for the Physics of Complex Systems, Nöthnitzer Str. 38, 01187 Dresden, Germany}
\maketitle

%\title{Towards a statistical physics of dating apps}
%\author{Fabrizio Olmeda \footnote{fabrizio@pks.mpg.de}} 
%\affil{Max Planck Institute for the Physics of Complex Systems, Nöthnitzer Str. 38, 01187 Dresden, Germany}
%\date{}
%\maketitle

%\begin{linenumbers}

\section*{Abstract}
Over the last ten years, a sharp rise in the number of dating apps has broadened the spectrum of how one can get in contact with new acquaintances. A common feature of such apps is a swipe enabling a user to decide whether to like or dislike another user. As is the case in real life, a user may be more or less popular, which implies the distribution of likes among different users is not trivial. In this paper we show how likes are distributed across users, based on different decision-making strategies and on different app settings. We apply theoretical methods originally developed in stochastic and coagulation processes to the investigation into the dynamics of dating app networks. More specifically, we show that whenever a dating app differentially displays different users with respect to their popularity in different models, users are split into two categories: a first category including users who have received most likes and a second category, referred to as a condensate, which in the long-term will be reduced to a small fraction of likes or to no likes at all. Finally, we explore different models based on a rating system of the users, known as Elo. These models will turn out to exhibit behaviour typical of gelating systems and non-trivial distribution of Elo rating.

%TC:endignore

\section*{Introduction}

There is huge variation amongst dating apps in terms of target audience, design, and popularity. However, most predicate on the common feature of letting the user decide to whom they are attracted to and want to start a conversation with. In many cases, they will suggest some users over others based on common interests, goals, etc... Being attracted to a particular user is translated into a user giving a like to another one or a dislike. When both users give each other a like, the pair is said to be a “match”. These two functions provide the basis of most dating apps. In most cases, a user cannot see all the active users at the same time, so as to decide to whom to give a like. Rather, the user is faced with one user at a time and decides whether to like or dislike sequentially. The statistics of matches and likes in this case is determined by the choices of the users and a network of possible connection emerges. Many questions arises from the dynamics on this networks, for example: how many people should we date before finding the right one? \cite{37Rule}, also known as the 37\%. rule. 

\par  The factors involved in the decision-making process of each individual user can be very different from one to another. Some users may be more interested mostly in the physical attractiveness, others on shared hobbies and passions, or a mixture of both. As there are many different factors that determine whether a user is keen to give or not a like to another user, the decision-making process is not trivial at all \cite{FacialAttractiveness,BluredFace}. As an example, in \cite{Distractor}, it has been shown that we may judge a person more or less attractive in comparison to other users we judged before. In \cite{Twin}, they asked different groups to rate pairs of twins between 1 to 7 and it was shown that there was a significant difference in the average rating of the twins, illustrating the intrinsic variability in perception of physical attractiveness. However, in case the distribution of rating does not show fat tails, meaning that the average “rating”, as well as its standard deviation, are well defined, then individual preferences will be smeared out on large numbers. In \cite{Twin}, the standard deviation was smaller by more than a factor of four with respect to the mean. In the following we will define $x_i$, the attractiveness of a particular user, as it would be perceived on average.

\par In this work, we explore the dynamics of likes in dating apps networks based on different decision making processes and possible different settings of the app. In particular, different users may be shown differently with respect to certain criteria, for example, how many likes they get on average, or the difference between likes and dislikes, which will likely determine their popularity, to which we will refer to as rank or rating. We will show that for different models, whenever an app sets a liked-based visibility, the users will be split into two group: a relative small group that will receive most of the likes and a bigger one, which we will refer to as a condensate, that in the long term, will receive few likes, and eventually none. In the last section, we will apply an Elo-based rating system \cite{ELO}, often used for sports such as chess and tennis, which rates  users in a non-trivial way with respect to the number of likes/dislikes received. We will show that even in this more complex scenario, the condensates still exists and in some particular cases the network will exhibit gelation. Everywhere, we will define a network of users and  we will not take into account gender.
\section*{Results}
\subsection*{Model definition}
We will now define a very general model for the dynamics of likes/dislikes/matches in dating apps. In the following sections we will focus on specific examples. For each user we define a variable $x_i$, which is the average perceived attractiveness of the user $i$. Average is meant as the average  of the attractiveness that each other user will assign to the user $i$. Attractiveness is a complex variable, that in general may depend on many different factors, but as in \cite{Twin}, we will just take it to be a scalar.
\par Moreover, a  dating app may decide to not show every user equally likely. In particular, they may define a certain criterion for which a user is shown more or less than other users. One possible criterion may be to show users with a probability that depends on the difference between the number of likes and dislikes received, i.e. ranking the users. We will refer to the rank of users $i$ as $c_i$. 
We define the following dynamics on the network of users (\textbf{Fig 1A}): \\
At each time step of the simulation, we randomly select a user $i$ and another user $j$ with Gillespie rules \cite{Gillespie} (\textbf{Methods}), based on the interaction Kernel $K(c_i,c_j)$. The pair is select with order. By order, we mean that we select the interaction $i \rightarrow j$ and not $j \rightarrow i$, as this is what typically happens in dating apps. 
The probability that the user $i$ gives a like to $j$, $P(J_{ij} =1)$,  is proportional to a function of the average perceived attractiveness of the users $f(x_i,x_j)$ and the probability of dislike is $P(J_{ij} =-1) =1 - P(J_{ij}=1)$. Time ($t$) is updated after each interaction event as $t \rightarrow t -  \frac{\log(u)}{\sum_j K(c_i,c_j)}$, with $u$ a uniform random number in $[0,1]$. 
\par We do not consider the history of the users, this is to say, that we do not consider whether a like/dislike was already given. In particular, typically a user may decide to remove the like, which we consider equivalent to a dislike or a user may change his/her mind toward another user if the app proposes that user again.  We  then extract another pair and repeat the procedure and a network of interactions emerges  (\textbf{Fig 1B}).
In case both users give a like to each other, this is called a ``match". A user can likewise remove a like and consequentially the match. In a statistical physics sense, we are looking at the dynamics of the couplings (likes/dislikes), rather than nodes (users). 
\par In the following, we will  consider two possible scenarios for the probability of giving a like: biased and non-biased. 
In the first case (biased),  the decision of a user $i$ will depend on the difference in attractiveness between him/herself ($x_i$) and the  attractiveness of proposed user $j$  by the app ($x_j$). For non-biased decisions, the probability of like will depend only on the attractiveness of $j$. 
We will take everywhere, unless specified otherwise, a fixed network, where the number of users is a constant.

\subsection*{Biased decisions}

In the case of biased decision making, a user $i$ will make a decision based on the difference between the attractiveness of him/herself and of user $j$ ($\delta_{ij} = x_i - x_j$). A very simple form for this interaction is taken to be:
\begin{equation}
P(J^t_{ij} = 1) = min(1,e^{-\beta |\delta_{i,j}| })  \, .
\end{equation}

$P(J^t_{ij} = 1)$ is the probability of user $i$ giving a like to  user $j$ at time $t$, or keeping a like in case the like is already present. The probability of removing a like is then $P(J^{t}_{ij} = -1) = 1-P(J^{t}_{ij} = 1) $. This probability is $1$ when user $i$ perceives user $j$ as more attractive than him/herself. In the other case, the like is given with a decreasing probability as the difference of attractiveness increases. $\beta$ is the average ``pickiness" or bias. When $\beta = 0$, every given user will like every other user regardless of their attractiveness. Instead, when $\beta \rightarrow \infty$ a user $i$ will never give a like to another user $j$, unless $j$ is more attractive than $i$. 
\par Initially, we will consider attractiveness to be a random variable, Gaussian distributed with a mean $\mu$ and variance $\sigma^2/2$. The difference between attractiveness ($\delta_{ij}$) of two Gaussian random variable is a random variable Gaussian distributed with mean $\mu = 0$ and variance $\sigma^2$, which we will set to one. The network and its dynamics are fully defined and we can then study long term statistics. In particular (\textbf{SI}), the probability distribution of the match probability  is,

\begin{equation}
\label{eq:matchmatch}
P(m) = \sqrt{\frac{2}{\pi}} \frac{1}{\beta m} e^{-\frac{\log(m)^2}{2\beta^2}}, m \in [0,1]
\end{equation}
 where $m$ is the probability of a match. $P(m)$  is a log-normal distribution , for high values of $\beta$  most of the pairs have low probability of a match, whilst the opposite is true for low values of $\beta$. We define $\beta_L$, the value of $\beta$ such that, if $\beta<\beta_L$ more than $50 \%$ of couples have more than $50\%$ probability of a match,
 \begin{equation}
\beta_L := \int_{0}^{0.5} \text{d}m P(m,\beta_L) = 0.5.
 \end{equation}
 
Numerically, $\beta_L \approx 0.83$.  Another statistically relevant quantity is the marginal probability of matches giving the attractiveness of a user. In other words, how many matches are you expected to have if you are attractive $x_i$?  The results (\textbf{SI}) are shown in \textbf{Fig2a, Fig2b} for two different distributions of attractiveness, respectively uniform and Gaussian. In both cases, attractive users are expected, on average, to get less likes than user whose attractiveness is closer to the mean value of the distribution. This apparently surprising effect occurs because users, who are perceived very attractive are likely to not give many likes and on top of that, they are on the tails of the distribution of attractiveness and thus represent the minority of users. Combining these results give them a low chance of having a match, whilst user, whose perceived attractiveness is close to the mean $\mu$, will be the one with most matches, but  receive less likes. The difference in matches percentage is smoothed  with a uniform distribution of attractiveness, as in this case, the very attractive users are not a minority. For networks with biased decision making, the dynamics of matches and likes can be quite different, which  in the following, we will prove to not be the case for non-biased decisions.

\subsection{Non-biased decisions}

In the case of non biased decision making, a user $i$ will make a decision solely based on the attractiveness of the user $j$. A very simple form for the probability of giving a like is,
\begin{equation}
P(J^{t}_{ij} = 1) = x_j^{\alpha} \, ,
\end{equation}

where $\alpha$ plays the same role of $\beta$ in the previous example. We made this particular choice for the sake of notation. We could have taken $P(J^{t}_{ij} = 1) = e^{x_j^{\alpha}}$ as before or other functions, but as it becomes clear later, the particular choice does not influence the dynamics of the models we will consider. In particular, $\alpha = 0$ means user giving likes to every other user and $\alpha \rightarrow \infty$ corresponds to no likes in the network. We will assume everywhere $\alpha>1$ and $x \in [0,1]$. In this very simple example,the probability of a match between users $i$ and $j$ is given by $p(m_{ij})= (x_ix_j)^{\alpha}$. 
\par Differing from the previous case, the average number of matches is a monotonically increasing function of $x_i$, meaning that the number of matches  of users with attractiveness $x_i$ is trivially proportional to itself: $\langle n(x_i)\rangle = x_i^\alpha E[x^\alpha]$,
where $E[\dots]$ is the expected value over the distribution of attractiveness. In the following, unless stated otherwise, we consider uniform distribution of attractiveness. The monotonic behaviour is not changed  by the specific choice of the distribution of attractiveness. In the uniform case, $\langle n(x_i)\rangle =x_i^{\alpha}/(1+\alpha)$. This result is quite trivial and is likely to be far from a real dating app network. In the next section we are going to model different restrictions for the possible ways in which users can give likes.

\subsection*{Visibility }

A dating app may decide not to show every user with equal probability.  We consider a certain criterion for which a user is shown more or less then other users . One possibility of such criterion may be to show users with a rate that depends on the difference between the number of likes and dislikes received, i.e. ranking the users.
Analytical insights regarding the number of matches in such conditions are hard to obtain, as they involve two-pair interactions. In this scenario, we can focus on the dynamics of likes as they will be a good statistics to infer the dynamics of matches as we will show later for different numerical simulations.
\par As the visibility is proportional to the number of likes/dislikes received, we may expect that that the probability of a match is proportional to the simple multiplication of these factors. This naive reasoning will lead to a probability of a match:  $p(m_{ij}) \sim (x_ix_j)^{\alpha}v_iv_j$, where $v_i$ is the visibility of a user $i$. 
We take visibility to be a function of the average number of likes ($k_1$) minus the dislikes ($k_2$) received $v = v(k_1-k_2)$. As we will show later, $k_1-k_2 \sim 1-2x_i^{\alpha}$, so that the argument of the visibility function has a sign change when $x_i^{\alpha}=1/2$ . Intuitively, we found a threshold in attractiveness $x_C = (1/2)^{1/\alpha}$, which will influence the long term behaviour of the system.
In particular, we will show, that likes and matches are concentrated into users that have a perceived attractiveness $x >x_C$, whilst other users will have very few to no matches. These phenomena resemble condensation phenomena in physics \cite{FreyCondensation,Bose-Einstein} or the wealth distribution in asset exchange models \cite{WealthDistribution,Marsili}, where in certain condition the resources are concentrated in few traders. 
\par The reason of the failure of the naive approach comes from taking a stationary perspective. Instead, we have  to consider the full dynamics, as this system will turn out to exhibit absorbing states \cite{FirstExit}. If  the visibility of a user goes to zero, that user will not be shown anymore and we will refer to the user as to be in a condensate.
This seems to be a very extreme case, which will  likely turn out not to be profitable for the app. We will consider a different case later, when the app still ensures a constant visibility. 
In order to derive analytical insight into the dynamics of the network, we propose mean field models, such that we consider users to be in a ``bath" of other users. This is to say that in the analytical models we will not consider the specific interaction between two users, rather we study the dynamics of likes/dislike received.
\par We define $n(k_1,k_2,c,x,t)$, the density of users that received $k_1$ likes, $k_2$ dislikes, are in a rank $c$ and are attractive $x$ at time $t$.  The mean-field dynamics of  $n(k_1,k_2,c,x,t)$ is given  by (for $c =1,\ldots, N-1$),
\begin{equation}
\label{eq:visibility}
\begin{split}
   & \partial_t n(k_1,k_2,c,x,t) = \omega  n(k_1-1,k_2,c-1,x,t)v(c-1)x^\alpha  \\&+  \omega n(k_1,k_2-1,c+1,x,t)v(c+1)(1-x^\alpha) - \omega n(k_1,k_2,c,x,t) v(c) \, ,
\end{split}
\end{equation}
and $\omega$ is the swiping rate,is set to one. We omit, for the sake of simplicity, the boundary equations at $c=0$ and $c=N$, but they are easily derivable. We consider the case in which, every time a user gets a like, his/her category is increased by one and decreased by one in case of a dislike. The first term on the r.h.s of Eq. \eqref{eq:visibility} takes into account that the density of users with $k_1$ likes increases whenever a user, who had  $k_1-1$ likes gets a like and the rank is increased by one as well. The second term accounts for dislikes and the third is the loss term due to getting a like or dislike. We take the visibility, which is the typical rate at which a user  is seen, to be  equal to the rank , $v(c) = c/N$. We will  make the time adimensional, by $\omega/N$ and  refer to it simply as $t$.
Upon defining a moment generating function, $G(z_1,z_2,z,x,t) = \sum_{k_1,k_2,c}z_1^{k_1}z_2^{k_2}z^c n(k_1,k_2,c,x,t)$, the average number of likes $\langle k_1(x,t) \rangle$ and the average category $\langle c(x,t) \rangle$  for users that are attractive $x$ at a time $t$ is given by (\textbf{SI}),
\begin{equation}
\label{eq:moments_visibility}
     \langle k_1(x,t) \rangle = c_0 x^{\alpha} \frac{e^{t(2x^{\alpha}-1)}-1}{2x^{\alpha}-1} \, ,  \, \langle c(x,t) \rangle = c_0 e^{t(2x^{\alpha}-1)} \, , \, x \neq x_C \, .
\end{equation}
$c_0$ is the initial rank. The average rank will increase for users such that $x>x_C$ and goes to zero for $x<x_C$. The number of likes will increase exponentially whenever $x>x_C$ and saturate exponentially for $x<x_C$. The time scale at which this saturation occurs is  set by: $\tau = \frac{1}{|2x^{\alpha}-1|}$. This time scale then sets the expected time before reaching the condensate. This simple exponential relationship is expected to hold only for a time such that $t < 1/\tau$ and for $x<x_C$. The reason for the latter inequality is in the boundary conditions for the rank $c$, as  Eq.\eqref{eq:moments_visibility} predict an exponential growth beyond $c=N$ for $x>x_C$.  

\par  In \textbf{Fig2C} (top,bottom), we compare Gillespie simulations \cite{Gillespie} of the exact agent based model (\textbf{Methods}) to analytical solutions from Eq.\eqref{eq:moments_visibility}. In \textbf{Fig2C} (top), we show the average category with respect to attractiveness at different times. As expected, there is a sharp transition at $x=x_C$, signalling condensation of likes and matches in few users.  In \textbf{Fig2C} (bottom), we compare the number of matches from the simulation to the number of likes. A linear regression is sufficient to fit the data, meaning that, as expected, the number of likes received are a good statistic for the matches in this model.
It is important to recall that the threshold is not only given by the app ranking the users, but by the choices of the users themselves. Large values of $\alpha$, means that a user is keen to reject almost every other user. 
Even though, on average, the category of users with $x>x_c$ increases over time, not every user for whom $x>x_C$ will escape from the condensate. As this is a first exit problem \cite{FirstExit}, the first moments will not be sufficient to answer the question.
\par In \textbf{Fig2C} (top,inset) we compare the dynamics of likes for users with $x<x_C$ ($x=0.3,\alpha=2$).  Numerical and theoretical solutions are in good agreement, even beyond the expected time-scale. As a remark, we defined likes received by a user as effectively to be likes plus the number of times the other users decide to keep the like. In \textbf{Fig2C} (top,inset) we compare the results for this definition of likes (green) to the correct one (purple). The difference between these definitions increases as we reach the condensation threshold and goes to zero for $x \ll x_C$. A theory beyond mean field is required to properly take the right definition of likes into account.

\par In this simple model, it is easy to write an equation for the dynamics of the ranks only,
\begin{equation}
\label{eq:birth_death}
    \partial_t n(c,x,t) =  x^\alpha(c-1)n(c-1,x,t) + (1-x^\alpha)(c+1)n(c+1,x,t) - cn(c,x,t) \, .
\end{equation}

Eq .\eqref{eq:birth_death} is a known birth-death process \cite{Kendall}, which after defining a generating function $G(z,x,0) = \sum_c z^c n(c,x,t)$, can be solved with the method of characteristic.
$n(c,x,t)$ is found by a series expansion of $G(z,x,t)$ \cite{CrawfordSuchard} and the number of users in the condensates is
\begin{equation}
    n(c=0,x,t) = \left[ \frac{(1-x^{\alpha}) - (1-x^{\alpha})e^{t(1-2x^{\alpha})}}{x^{\alpha}- (1-x^{\alpha})e^{t (1-2x^{\alpha})}}\right]^{c_0} \, .
\end{equation}

Whenever $x<x_C$ in the asymptotic limit  the steady state value for the number of users in the condensate is : \\
$n_s = \lim_{t\rightarrow \infty} n(0,x<x_C,t) = 1$, whilst $ n_s = \lim_{t\rightarrow \infty}n(0,x>x_C,t) = (\frac{1-x^{\alpha}}{x^{\alpha}})^N$. This is quite surprising as even a user who is perceived more attractive than the condensation threshold, may still end up in the condensates with a finite probability. This probability monotonically decreases with the initial value of the rank.
This model gives a very strict criterion and will not ensure any guarantees of matches to the users, meaning that a user will most likely choose some other app. In the next section, we analyse different possibilities for a dating app to prevent condensation, but keeping a rank based approach.

\subsection*{Secured visibility}
We consider two possible scenarios in which an app may secure a constant visibility to avoid condensation. In the first case, we consider that users will still been shown with a given rate no matter their rank.
The dynamics of $n(k_1,k_2,c,x,t)$ is given by,
\begin{equation}
\label{eq:securevisibility}
\begin{split}
   & \partial_t n(k_1,k_2,c,x,t) =   \gamma n(k_1-1,k_2,c-1,x,t)v(c-1)x^\alpha  \\&+ \gamma  \left[n(k_1,k_2-1,c+1,x,t)v(c+1)(1-x^\alpha) - n(k_1,k_2,c,x,t) v(c)\right]\\
   &+(1-\gamma)\left[n(k_1-1,k_2,c-1,x,t)x^\alpha + n(k_1,k_2-1,c+1,x,t)(1-x^\alpha) - n(k_1,k_2,c,x,t) \right] \, ,
\end{split}
\end{equation}
where $\gamma$ sets the relative weight of the constant visibility and the like-based visibility.  In particular, terms in Eq.\eqref{eq:securevisibility} that carry a $1-\gamma$ factor do not involve any visibility criterion, such that the rate to be seen is equal to $1-\gamma$ for every user.
In the following, for every model, we rescale all rates to avoid an explicit dependence on $N$ and we will not discuss it further as we will be mostly interested in long-term statistics.
As in the previous section we can simplify this equation by integrating out $k_1,k_2$ and study the dynamics of the ranks only,
\begin{equation}
\begin{split}
  \partial_t n(c,x,t) = & \gamma \left[x^\alpha (c-1)n(c-1,x,t) + (1-x^\alpha)(c+1)n(c+1,x,t) - cn(c,x,t)\right]  \\ 
 &+ (1- \gamma)\left[x^\alpha n(c-1,x,t) + (1-x^\alpha)n(c+1,x,t) -  n(x,c,t)\right] \, .
\end{split}
\end{equation}
 The asymptotic behaviour of $n(c=0,x<x_C,t)$ is given by \cite{TransientBirthDeath},
\begin{equation}
\label{eq:Zheng}
    n(c=0,x<x_C,t\rightarrow \infty) = \frac{\gamma}{1-\gamma} \left[ \int_0^1 ds \left( 1- \frac{x^{\alpha}}{1-x^{\alpha}}s \right)^{\frac{\gamma-1}{\gamma}} \left(1-s\right)^{\frac{1-2\gamma}{\gamma}}\right]^{-1} \, ,
\end{equation}

in which the limit $\gamma,x^{\alpha} \rightarrow 0$ scales as $ n_{s} = n(0,x\rightarrow 0 ,t\rightarrow \infty) \sim (1-x^{\alpha})+\gamma$. This implies that the particular policy of the app and the probability for which a user decides to give a like contribute equally  to the number of likes received. The number of likes received can be estimated from Eq.\eqref{eq:securevisibility} and its approximate asymptotic limit in case $x<x_C$, $\lim_{t\rightarrow \infty} \langle k_1(x,t) \rangle = (1-\gamma)x^{\alpha}t$. As seen in numerical simulation  \textbf{Fig2D} (top) , there is still a sharp gap in the number of likes at $x=x_C$ and the average category which may be closed only when $\gamma =1$. As the previous model, the likes are still a good statistic for the number of matches,  \textbf{Fig2D} (bottom). 
\par In order to compare theoretical results with simulations of the agent-based model, we need to consider that even though the number of likes for users such that $x<x_C$ increases linearly,  it is always a negligible fraction of the total amount of likes given in the network: $\lim_{t \rightarrow \infty} \frac{\int_0^{x_c} dx \langle k_1(x,t) \rangle }{\int_{0}^1 dx \langle k_1(x,t) \rangle} = 0$. This way of securing a constant visibility,  though increasing the number of likes for users such that $x<x_C$, will never break the condensation mechanism. Moreover, in comparison with the previous model, the average rank decays faster to zero \textbf{Fig2C}, \textbf{Fig2D} (top). This apparently surprising behaviour is explained by intuitively noticing that being more visible outside of the condensate increases the chance of being rated lowered for users with $x<x_C$.
\par Another possibility in which the app may secure a constant visibility is a ``push" strategy.  As an example, we consider that with a certain rate ($1-\gamma$), the app increases the rank of every user to avoid condensation. The dynamics of the density of users in a given rank is described by the mean-field master equation,
\begin{equation}
\label{eq:push}
\begin{split}
  \partial_t n(c,x,t) = & \gamma \left[x^\alpha (c-1)n(c-1,x,t) + (1-x^\alpha)(c+1)n(c+1,x,t) - cn(c,x,t)\right]  \\ 
 &+ (1- \gamma)\left[n(c-r,x,t) -  n(x,c,t)\right] \, .
\end{split}
\end{equation}

We set again $\gamma$ as the relative contribution in order to compare the result with the previous case. $r$ is the amount of ranks that the app decide to push users up. Eq.\eqref{eq:push} has a similar form of master equation for gene expression with translational burst \cite{ShahrezaeiSwain, MuglerWalczak}, which in this case and for $r =1$ can be solved exactly. The solution can be found again by the method of characteristics (\textbf{SI}). The density of users in the condensate and the average rank, for $x<x_C$, scales as,

\begin{equation}
\label{eq:push_saturation}
  n(0,x<x_c,t\rightarrow \infty) = \left[1- \frac{x^{\alpha}}{1-x^{\alpha}}\right]^{\frac{1-\gamma}{\gamma x^{\alpha}}} \, \langle c(x<x_C,t \rightarrow \infty) \rangle =  \frac{1-\gamma}{\gamma(1 -2   x^{\alpha})}
\end{equation}
with $\lim_{t\rightarrow \infty} n(0,x>x_c,t) = 1$.  The average rank does not decay exponentially to zero as before, but rather saturates to a constant value, assuring always a non zero visibility to every user. However, even in this scenario, the number of likes received by the users for which $x<x_C$ compared is still negligible compared with the total number of likes given in the network: $\lim_{t \rightarrow \infty} \frac{\int_0^{x_C} dx \langle c(x,t) \rangle }{\int_{0}^1 dx \langle c(x,t) \rangle} = 0$ . 
\par In \textbf{Fig2E} (top) we plot results of simulation of the corresponding agent based model with the theoretical bound. Even in this model, likes are sufficient to predict the matches  \textbf{Fig2E} (bottom). The reason why this strategy is more profitable for users than the previous one for $x>x_C$ is straightforward. In the ``push" strategy, a user may climb some ranks and eventually rise far enough from the condensate, such that  it will take a long time to return there. In the case of a secured visibility, it is only the probability of receiving a like that can bring a user out of the condensate, which is clearly not profitable for users such that $x \ll x_C$. 
\par So far we only considered very simple models in order to understand how relevant quantities, such as likes or matches, change in time and we identify a clear threshold for condensation. In the next paragraph, we will present more realistic models and compare them with the minimal models that we have analysed.

\subsection*{Elo Rating System}
In some sports like chess or tennis, players are rated based on an Elo score \cite{ELO}. A Elo rating system assigns to players a certain score, which changes with respect to the outcomes of individual or tournament games. Here, we consider the scenario in which dating apps decide to adopt a similar rating system and we are going to draw conclusions on the number of likes received on average by each user and the effect that this ranking system might have on condensation. A general feature of Elo rating system is that a user, who receives a like from another user in a higher rank, will climb ranks directly proportional to the difference between the rank of the users. The opposite holds true in case of a dislike.
\par We define $n(c,x,t)$ as the density of users in the rank $c$ who are attractive $x$, whose dynamics is given by,

\begin{equation}
\begin{split}
\label{eq:ELO}
 & \partial_t n(c,x,t) = \sum_{w=0}^{\infty}v(w)n(w,x,t) \sum_{j=0}^{\infty}  \left[ \int_0^1 \text{d}y P(y) n(j,y,t)  \left( x^{\alpha} K^+(w,j,c) + (1-x^{\alpha})K^-(w,j,c)\right) \right] \\
 &-n(c,x,t)v(c) \left[ \sum_{w,j=0}^{\infty} \int_0^1 \text{d}y P(y) n(j,y,t)  \left( x^{\alpha} K^+(w,j,c) + (1-x^{\alpha})K^-(w,j,c)\right) \right]\,
\end{split}
\end{equation}
with $P(y)$ the distribution of attractiveness.
Differently from the previous models, the rate at which the the density of users in the rank $c$ increases or decreases, after a like or dislike, is proportional to the product of users' densities, rather than linear. The quadratic term then appears because we need to take into account the rank of the users that give  a like or dislike. In Eq. \eqref{eq:ELO}, $K(w,j,c)$ is the rate at which the app proposes a user with Elo $j$  to a user with Elo $w$ and $v(w)$ is the visibility rate as before. $K^{\pm}(w,j,c)$ stands for like/dislike respectively. The extra variable $c$ in the function $K(w,j,c)$ indicates that the final Elo of a user, with initial Elo $w$, is updated to $c$, after receiving a like from a user with Elo $j$. The same reasoning applies to the case of a dislike. We will consider the case that a like from a user in a lower category or a dislike from a user in an higher category makes way less impact than a like or dislike from a user in a higher or lower category, respectively.  This will give some constraint on the function $K(w,j,c)$.  In the next sections we are going to present detailed numerical simulations of a version of Eq.\eqref{eq:ELO} based on Elo rating systems. 

\subsection*{Product Kernel}

As a first simple model, we introduce a product Kernel, such that  $K^{\pm}(w,j,c)= \delta(w,c\mp 1) (j+1)$ in Eq.\eqref{eq:ELO}. Intuitively, a product Kernel increases the chance of users, who are in higher ranks, to actively use the app. Users, who received more likes and so matches, are likely happier with the dating app and they use it with higher frequency compared to other users. In this case, the dynamics of the density of users is described by ($v(w) = w$ as previously),

\begin{equation}
\label{eq:linear_elo}
\begin{split}
 & \partial_t n(c,x,t) = x^{\alpha}(c-1)n(c-1,x,t) \int_0^1 \text{d}y P(y) \sum_{j=0}^{\infty} (j+1) n(j,y,t) 
 \\ & + (1-x^{\alpha}) (c+1)n(c+1,x,t)  \int_0^1 \text{d}y P(y) \sum_{j=0}^{\infty} (j+1) n(j,y,t)- cn(c,x,t)   \int_0^1 \text{d}y P(y) \sum_{j=0}^{\infty} (j+1) n(j,y,t)   \, .
\end{split}
\end{equation}

 Upon rescaling time as $\tau =  \int_0^{t} dt' \int_0^{1} dx   \left[\partial_zG(z,x,t) + G(z,x,t)\right]_{z=1}$, with $G(z,x,t) = \sum_c z^c n(c,x,t)$, the number of users in a condensate is (\textbf{SI}),
\begin{equation}
\label{eq:main_gel}
n(c=0,x,\tau) = \left[ \frac{(1-x^{\alpha}) - (1-x^{\alpha})e^{\tau(1-2x^{\alpha})}}{ x^{\alpha} -(1-x^{\alpha})e^{\tau (1-2x^{\alpha})}}\right]^{c_0} \, 
\end{equation}

with $ \tau \approx \log[\frac{1}{(1+c_0f_{\alpha})e^{-t} -c_0f_{\alpha}}]$. Differently from the previous models, Eq.\eqref{eq:main_gel} is not properly defined at times $t>\log(\frac{1+c_0f_{\alpha}}{c_0f_{\alpha}}) = t_g$, signalling a drastic change of the Elo-ranking distribution. We then compute second and higher order moments of Eq.\eqref{eq:linear_elo} and we found that they all diverge at $t =  t_g$ (\textbf{SI}), which is a signature of a gelation transition \cite{KineticView}.

\par Gelation occurs in typical aggregation processes when, at finite times, one cluster has a non vanishing fraction of the total mass. In models of dating apps, a giant cluster or gel, is represented by few users, who have very high rank (mass) compared to the majority of users. The Elo score of the ``gel"  increases over time at the expense of users with lower Elo scores. Differently from gelating system, the first order moment diverges in our system, as the total mass, is not a conserved quantity. In \textbf{Fig3B} we confirm this theoretical prediction by showing how the Elo distribution evolves in time. We notice that the  distribution flattens over time with a high peak around zero and a broad tail that moves dynamically to higher Elo values, as expected in gelating system \cite{Gelation}.  As in the previous models, users with $x>x_C$, will experience an indefinite increase of their rank, which is unlikely to be realist. In the last section we will derive a general model for a chess-based Elo rating system, which will fix this unwanted feature.

\subsection*{Chess-based Elo}

We initially relax the assumption of unitary transition between ranks. This means that a user is able to jump more than one rank whenever receives a like from a user in a higher rank. In particular, we will consider a linear chess-based Elo rating system \cite{ELO}). In this case, whenever a user in a rank $w$ receives a like/dislike from a user in a rank $j$, his/her new rank $c$ is,

\begin{equation}
    c = w \pm \frac{W}{2} + \frac{W}{4C}(j-w) \, ,
\end{equation}
With a positive sign in the case of a like and negative for a dislike (in the book of Elo $W$ is $K$, here we use $K$ for the Kernel).
We define $W/4C = \zeta$ and $W/2 = \zeta \gamma$. An Elo ranking system, then provides a way to rank users which is not trivially related to the number of likes. When sports are provided with such rating system, it yields a way to decide whether two different players will compete. Indeed, it is quite unlikely to observe an Elo rated match between the world chess champion and a club player or even a grand master with a substantial Elo difference.
\par This last constraint is encoded in the interaction kernel $K(w,j,c)$. In \textbf{Fig3A} we outline the difference between chess-based Elo and the Elo in a model for dating apps. Eq.\eqref{eq:ELO} with the Elo chess-based representation for the dynamics of ranks  becomes,

\begin{equation}
\label{eq:general_elo_chess}
\begin{split}
    & \partial_t n(c,x,t) = x^{\alpha}\sum_{w(1-\zeta) + \zeta j +\zeta \gamma=c} \int_0^1 \text{d}y P(y) v(w)n(w,x,t) n(j,y,t)K(w,j) + (1-x^{\alpha}) \\
    &\sum_{w(1-\zeta) + \zeta j - \zeta \gamma=c}\int_0^1 \text{d}y P(y) K(w,j)v(w)n(w,x,t) n(j,y,t) - n(c,x,t)v(c)\left[\int_0^1 \text{d}y P(y) \sum_{j} n(j,y,t)K(j,c)\right] \, .
\end{split}
\end{equation}

The product of the visibility $v(c)$ and the kernel results in a non trivial interaction between users  as the ratio of the probability of interaction between users with Elo $c$ and $c'$ is $v(c)/v(c')$, which is independent of the Kernel. Even in the case of the possibility of non-unitary jumps between ranks, high jumps are  strongly minimized by the users Elo. In the following, we will keep as in the previous section $v(c) = c$ and a general form of the kernel will be taken as, $K(c,c') = K(c-c') = \frac{1}{1+(c'-c)^2}$.  A Kernel defined in this way maximizes interactions between users with similar Elo and minimizes it, in case of high Elo difference. We ran Gillespie simulations of Eq.\eqref{eq:general_elo_chess} and we require $c$ to be bounded from below, but not above, as in the previous sections. We avoided the boundary at $c=N$, as Elo score for users with $x>x_C$, will not grow indefinitely due to the two-body interaction term. This term thus, as expected, stabilizes the ratings. 
\par In \textbf{Fig3C} we show that the condensate has not disappeared.  There is a smoother transition at the condensation threshold for the steady state average rank with respect to the attractiveness. Moreover, the number of likes received is more evenly distributed among the users (\textbf{Fig3D}), similar to the ``push" algorithm. In \textbf{Fig3E} we show  the distribution of Elo independently on attractiveness. Interestingly, the distribution flattens initially, and becomes bimodal in the long-term limit. The higher mode is broad, \textbf{Fig3E} (inset), signalling what we expected from studying the average rank: the Elo ratings are more proportionally distributed among the users with $x>x_C$. At the same time, Elo ratings are not evenly distributed  when we consider the totality of users as the condensate is still present. This particular chess-based ranking system turns out to not be more advantageous for users with $x<x_C$ than the ``push" algorithm in the previous section.
\par To conclude, this chess-based Elo ranking system seems the most realistic among the previously analysed model as the ratings reach a stationary distribution. In future works, it will be a reasonable starting point to analyse, for example, the effect of users installing and uninstalling the app.

\section*{Discussion}

In this work, we studied the dynamics of likes and popularity (rank) in possible models for dating apps. Upon mapping the dynamics on this network to typical problems in stochastic processes we were able to show how the rank and number  of interactions change in time. In particular, we analysed two possible ways in which users may decide whether to give or not a like to another user.  In the first case, users decide based on the difference between attractiveness between themselves and the one proposed by the app. In this scenario, we found that the statistics of likes and matches are quite different as users who are not the most popular, i.e. received less likes than other users, may have more matches. Although it may appear surprising, this phenomenon appears due to the distribution of average perceived attractiveness combined with the strict choice of the users. 
\par As a second class of models, we consider that  users decide solely on the other users attractiveness. This form of decision-making leads to simple percentage of likes and matches only in the case where all of the users are equally visible. Instead, when users are shown differently based on their popularity, there might be situations in which a finite fraction of the users will receive from few to no likes at all. This phenomena, typical of condensation mechanisms, will set a threshold in attractiveness, below which users enter into the condensate. We found, by exploring different possible features of the app, that the condensation mechanism is hard to break, as it will always have a strong component related to the users' swiping strategy. However, we found that whenever the app decides to increase with a certain frequency the rank of every user, the average rank and so the visibility will saturate to a constant value, smearing out the condensation phenomena.
\par In the last section, we  analyse models for dating apps in which users are ranked in a similar way as done in sports, such as chess or tennis. In such models,  we found typical features of gelating systems. In particular,  users will either end up in the condensates or they will experience an increase of the rank over time. On top, gelation leads to the accumulation of likes and rating to very few users. Finally, we use a chess-based Elo rating system, which is expected to  prevent gelation. We found that, even in this scenario, the condensation mechanism does not disappear, but ranks and likes are more equally distributed between users above the condensation threshold.
\par Our approach, based on a mean-field mapping, even though it is expected to give no accurate statistics over longer times, still catches the qualitative features of the analysed model. This approach makes it easier to study more complicated models, such as models on bipartite networks. As an example, bipartite networks arise when a dating app asks users to identify themselves with respect to gender. Our mean-field approach can be straightforwardly applied to bipartite networks as well, as it relies only on the dynamics of users in a ``bath" of other users belonging to the other networks. In particular, if the two networks have different average values of the ``pickiness" $\alpha$ or $\beta$, we can just replace the particular one in all our previous models, except the chess-based Elo. In this model, it is necessary to rescale ratings between the networks, to make them comparable. It will be interesting to better analyze this effect in other studies.
\par  Unfortunately, to our knowledge, there is no data available on the statistics of likes, matches and possibly ranks. It is then hard to point out which of these models better explain dating app algorithms. In the future, it will be interesting to explore different possible models with like based visibility to find  one which may smear out the condensation mechanisms outlined in this paper.  As evident from the models analysed here, the best way to avoid condensation is not to have visibility criteria.

\section*{Acknowledgments}
I thank F. Rost, F. Patacconi, F. Jülicher, Y. Qinghao, M. Popovic and S. Rulands for a critical reading of the manuscript. I thank L.Duso and all the members of the Rulands lab for helpful discussions.

\section*{Competing interests}
The author declares no competing interests.

\medskip
%\printbibliography
\bibliography{datingapps}

\newpage
\clearpage
\pagestyle{empty}

%TC:ignore

\section*{Figure legends}
\textbf{Figure 1}\\
\textbf{A}:  Possible interaction between a pair of users.  The two outcomes (likes or dislike) are determined by users' visibility $K(c_i,c_j)$ and the like probability $f(x_i,x_j)$. 
\\
\textbf{B}: Example of a possible configuration of the network. Users have a an averaged perceived attractiveness and they are ranked by the app. The decisions are determined by these two factors. As in the example, higher ranked users are more visible. 
\\
 \textbf{Figure 2}\\
\textbf{A}: Distribution of the average probability of matches as a function of the parameter $\beta$ (``pickiness") and the attractiveness. The distribution of attractiveness is Gaussian $N(0.5,0.1)$.
\\
\textbf{B}: Distribution of the average probability of matches  as a function of the parameter $\beta$ (``pickiness") and the attractiveness. The distribution of attractiveness is uniformly distributed in $[0,1]$.
\\
\textbf{C}: (top) Average normalized rank (with respect to the maximum) for unitary transition between ranks without any secured visibility from the app and  condensation threshold $x_C=(1/2)^{1 / \alpha}$ (dashed black line).(inset) number of effective likes (purple),likes(green) and theoretical prediction (dashed black). Time is in unit of swiping rate per number of users, such that, time equal to 1, is the average time between two swipes for a single user. (bottom) A linear regression between matches and likes is in agreement the simulated data (dots). 
\\
\textbf{D}: (top) Average normalized rank (with respect to the maximum) for unitary transition between ranks with secured visibility of the app $\gamma = 0.9$ and  condensation threshold $x_C=(1/2)^{1 / \alpha}$ (dashed black line). (bottom) A linear regression between matches and likes is in agreement the simulated data (dots).
\\
\textbf{E}: (top) Average normalized rank (with respect to the maximum) for unitary transition between ranks with ``push" strategy and asymptotic limit of the average rank from Eq.\eqref{eq:push_saturation} (dashed black line). (bottom) A linear regression between matches and likes is in agreement the simulated data (dots).
\\
\textbf{Figure 3}\\
\textbf{A}: Schematic illustration of the application of ELO rating system in chess compared with models  of dating apps.
\\
\textbf{B}: Distribution of ranks for the product Kernel $K(c,c') = c(c'+1)$.
\\
\textbf{C}: Average Elo for non unitary transition between ranks based on ELO-like ranking system.
\\
\textbf{D}: Average likes (rescaled by the maximum) for non unitary transition between ranks based on Elo-like ranking system.
\\
\textbf{E}: Distribution of ranks for $v(c) = c$, $K(c,c') = 1/(1+(c-c')^2)$, $\beta\gamma = 16$, $\beta = 0.02$. This choice are in accordance with the standard chess Elo rating system.(inset) Magnification of the higher mode of the distribution with logarithmic scale for Elo ratings. 
\newpage

\begin{figure}[H]
\centering
    \includegraphics[width =  \textwidth]{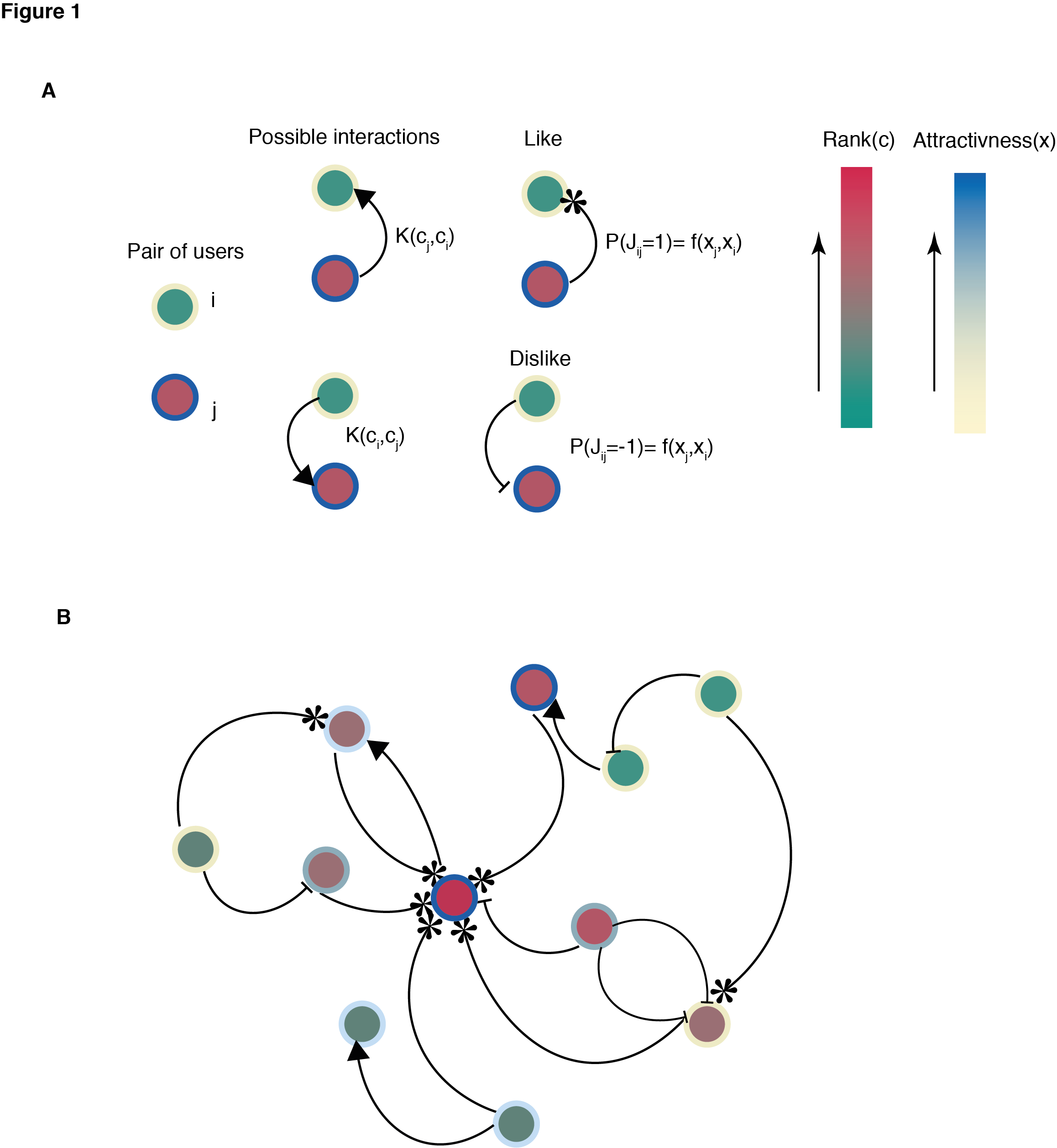}
\end{figure}
\begin{figure}[H]
\centering
    \includegraphics[width =  \textwidth]{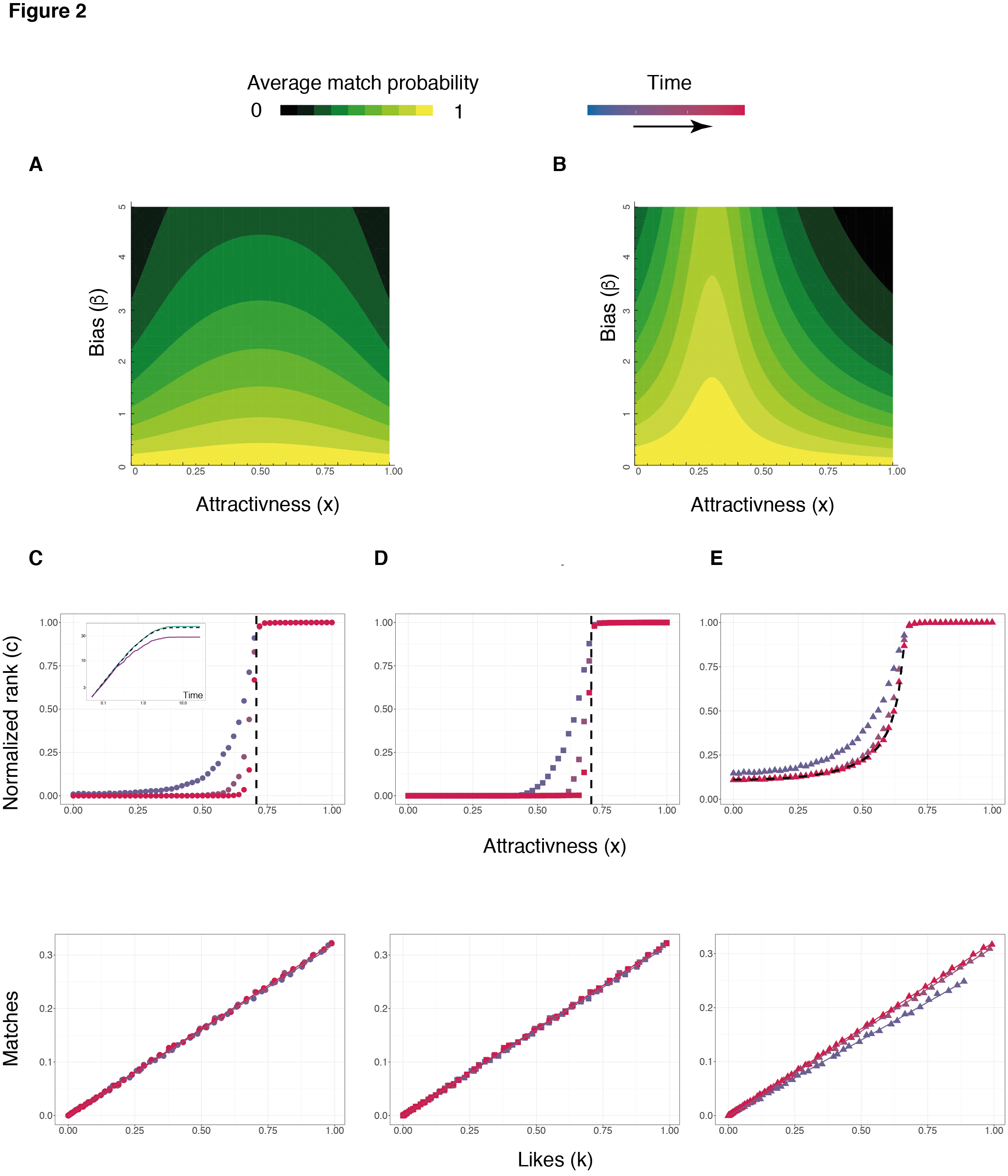}
\end{figure}
\begin{figure}[H]
\centering
    \includegraphics[width = \textwidth]{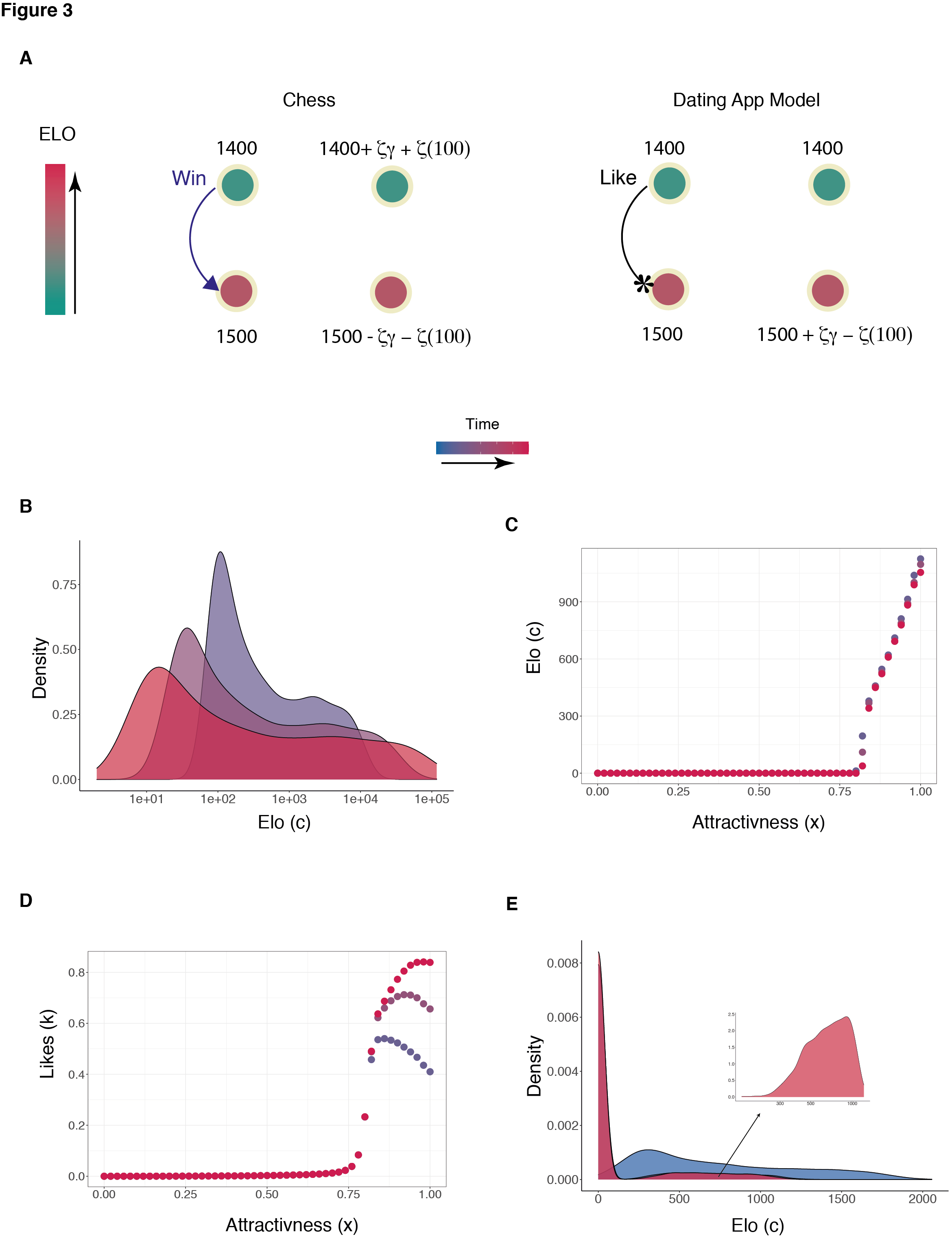}
\end{figure}

\newpage

\section*{Methods}
In this section we give numerical details for the Gillespie agent-based model simulations. In all simulations we fix the number of users to be $N = 1000$, we let the system evolve for $1e7$ time steps and the initial ranks are set to $c_0= N$ for every user. In case of Elo based ranking system, we initialize the Elo to $c_0 = 1000$. In the chess-based Elo rating system, we set $N = 1e4$ and $1e8$  time steps, to resolve the higher mode of the Elo rating distribution. We outline the numerical procedure for the model described by Eq \eqref{eq:visibility}, as it is general and the application to the other models is straightforward. We first initialise the network with $c_i = c_0, J_{ij} = 0, \forall i,j $,  where $c_i$ is the rank of a user $i$ and $\hat{J}$ is the matrix of likes. At every time step we select a user $i$. We then construct a vector of rates $\Vec{\lambda}$, which component $\lambda_i$, are the rates at which the user $i$ sees user $j$. For the given model, this rate is $\lambda_j = \gamma (c_j/N) + (1-\gamma)$ and $\lambda_i = 0$. We then choose randomly a user $k$ with a probability $\lambda_k/\sum_k \lambda_k$. With a probability $x_j^{\alpha} $ the rank of user $j$ is increased by one, if less than $N$ and zero otherwise. $J_{i,j}$ is set to one. With a probability $(1-x^{\alpha})$ the rank is decreased by one if $c_j>0$ and unchanged otherwise. $J_{i,j}$ is set to zero.  The number of likes received by user $j$ is $n_j = \sum_i J_{ij}$. 
\par For the Elo rating system, $\lambda_j = \frac{c_j}{1+(c_j-c_i)^2}$. After the pair is selected, with a probability $x^{\alpha}c$  the rank of user $j$ is updated to $c_j \rightarrow c_j + \beta \gamma + \beta(c_i - c_j) $ and $J_{i,j}$ is set to one and with a probability $(1-x^{\alpha})$,  the rank of the user $j$ is updated to $c_j \rightarrow c_j - \beta \gamma + \beta(c_i - c_j) $ and $J_{i,j}$ is set to zero.
Likes are computed as in the previous case.
For all the simulations, unless stated otherwise, attractiveness is uniformly distributed in $[0,1]$ and we computed average ranks, likes, matches for 50 linearly spaced bins of attractiveness.

\newpage
\clearpage
\pagestyle{empty}

\section*{Supplementary Information}
\renewcommand{\theequation}{S.\arabic{equation}}
\setcounter{equation}{0}
\subsection*{Distribution of matches for biased decision-making}
As the probability of a match $p(m_{i,j})$ is given by the product of the probability of giving likes: $p(m_{i,j}) = e^{-\beta| \delta_{ij}|}$,  the cumulative distribution is.
\begin{equation}
    F(p(m_{i,j}) \leq s) = F(|\delta_{ij}| \leq \log{t}/\beta) = \int_{-\log{s}/\beta}^{\log{s}/\beta} P(\delta_{ij}) d\delta_{ij} \,
\end{equation}
where $P(\delta_{ij})$ is the distribution of the difference of attractiveness. The probability distribution is then $P(s) = \partial_s \int_{-\log{s}/\beta}^{\log{s}/\beta} P(\delta_{ij}) d\delta_{ij}$. Upon taking $\delta_{ij}$ to be Gaussian distributed with zero mean and unitary variance, and after normalizing for $0<s<1$, we arrive to Eq. \eqref{eq:matchmatch}. To study the marginal probability of a match ($p(m_i)$) for a user attractive $x_i$ , we introduce  the cumulative distribution
\begin{equation}
      F_m(p(m_{i}) \leq s) =  \int_{-\log{s}/\beta +x_i}^{\log{s}/\beta + x_i} P(x_j) dx_j \, ,
\end{equation}
where the integration is performed only on $x_j$ and $P(x_j)$ is the distribution of attractiveness. The marginal probability is  given by $P(s) =\partial_s F_m(s)$, with $s$ replacing $m_i$ as before. The average probability of a match is: $\langle n \rangle = \int_0^1 ds \,  sP(s)$.
In case of uniform distribution the average probability of a match  for a user attractive $x_i$ is simply given by,
\begin{equation}
    \langle n \rangle  = \int_{x_i}^1 dx_j e^{-\beta(x_j-x_i)} + \int_0^{x_i} dx_j e^{\beta(x_j-x_i)} = \frac{1}{\beta}\left[2-e^{\beta(x_i-1)} + e^{-\beta x_i} \right] \, .
\end{equation}

\subsection*{Moment generating function }
Upon multiplying the r.h.s and l.h.s of Eq \eqref{eq:visibility} by $z_1^{k_1} z_2^{k_2}z^c$ and summing over $k_1,k_2,c$, Eq \eqref{eq:visibility} is expressed in terms of the generating function,
\begin{equation}
\label{eq:supp_generating}
\partial_t G(z_1,z_2,z,x,t) = \left[x^{\alpha} z_1 z^2 + (1-x^{\alpha}) z_2  - z \right]\partial_z G(z_1,z_2,z,x,t) \, .
\end{equation}

Moments are found by differentiation of the generating function,
\begin{equation}
\begin{split}
& \langle k_1(x,t) \rangle = \sum_{k_1,k_2,c} k_1 n(k_1,k_2,c,t)  = \left[ \partial_{z_1}G(z_1,z_2,z,x,t)\right]_{z_1=z_2=z=1}\\
& \langle k_2(x,t) \rangle = \sum_{k_1,k_2,c} k_2 n(k_1,k_2,c,t)  = \left[ \partial_{z_2}G(z_1,z_2,z,x,t)\right]_{z_1=z_2=z=1}\\
& \langle c(x,t) \rangle = \sum_{k_1,k_2,c} c n(k_1,k_2,c,t)  = \left[ \partial_{z}G(z_1,z_2,z,x,t)\right]_{z_1=z_2=z=1} \, .
\end{split}
\end{equation}
The equation for the dynamics of the moments  from Eq.\eqref{eq:supp_generating} are,
\begin{equation}
\begin{split}
& \partial_t \langle k_1(x,t) \rangle = x^{\alpha} \langle c(x,t) \rangle\\
& \partial_t \langle k_2(x,t) \rangle = (1-x^{\alpha}) \langle c(x,t) \rangle\\
& \partial_t\langle c(x,t) \rangle = (2x^{\alpha}-1)\langle c(x,t) \rangle \, .
\end{split}
\end{equation}
Upon solving the system of equations for $ \langle c(x,t) \rangle$ and plugging the solution into the other moments we arrive to Eq. \eqref{eq:moments_visibility}

\subsection*{Method of characteristics}
Upon defining the generating function $G(z,x,t) = \sum_c z^c n(c,x,t)$, Eq. \eqref{eq:push} is simplified to,
\begin{equation}
\label{eq:push_generating}
\partial_t G(z,x,t) = \gamma \left[z^2x^{\alpha} +(1-x^{\alpha}) -z \right]\partial_z G(z,x,t) + \left(1 - \gamma\right) \left[z^r -1\right]G(z,x,t) \, .
\end{equation}

Setting a characteristic $s$ starting at time $t = 0$ and for some initial value $z(s,t=0) = z_0$, Eq. \eqref{eq:push_generating} becomes,

\begin{equation}
\begin{split}
& \frac{dt}{ds} = 1\\
&\frac{dz(s)}{z(s)^2x^{\alpha} +(1-x^{\alpha}) -z} = ds\\
& \partial_s G(z(s),x) = (1-\gamma)\left[z(s)^r -1\right]G(z(s),x)  \, .
\end{split}
\end{equation}
From the first equation we find $t = s$, and so $z(s) = z(t)$. \\
From the second equation,
\begin{equation}
    z(t) = \frac{e^{(2x^{\alpha}-1)\gamma t}(1-x^{\alpha})(1-z_0) - (1-x^{\alpha})(1+z_0)}{e^{(2x^{\alpha}-1)\gamma t}x^{\alpha}(1-z_0) - (1-x^{\alpha})(1+z_0)} \, .
\end{equation}
And putting the results into the third equation,
\begin{equation}
G(z,x,t) = G(z_0,x) e^{(1-\gamma)\int dt \left[z^r(t)-1\right]} \, .
\end{equation}
If $G(z,x,t=0) = z_0 ^{c_0} $ then,

\begin{equation}
G(z,x,t)= \left[\frac{(1-x^{\alpha})(1-z) -e^{-(2x^{\alpha}-1)\gamma t} (1-x^{\alpha}(1+z))}{x^{\alpha}(1-z) -e^{-(2x^{\alpha}-1)\gamma t} (1-x^{\alpha}(1+z))}\right]^{c_0}e^{(1-\gamma)\int dt \left[z^r(t)-1\right]} \, .
\end{equation}
 If $r =1$, the solution is given by,
\begin{equation}
    G(z,x,t) = \left[\gamma (2x^{\alpha}-1)\right]^{\frac{1-\gamma}{\gamma x^{\alpha}}} \gamma^{-\frac{(1-\gamma)}{x^{\alpha}\gamma}}\frac{\left[(1-x^{\alpha})(e^{t(2x^{\alpha}-1)}-1) - z((1-x^{\alpha})e^{t(2x^{\alpha}-1)} -x^{\alpha})\right]^{c_0}}{\left[(1-x^{\alpha})(e^{t(2x^{\alpha}-1)}-1) - zx^{\alpha}(e^{t(2x^{\alpha}-1)}-1)\right]^{c_0 + \frac{1-\gamma}{\gamma x^{\alpha}}}}.
\end{equation}
The asymptotic distribution for the fraction of users in the condensate is then given by (for $x<x_C$),
\begin{equation}
  n(c=0,x,t\rightarrow \infty) = \lim_{t \rightarrow \infty} G(z=0,x,t)   = \left[1- \frac{x^{\alpha}}{1-x^{\alpha}}\right]^{\frac{1-\gamma}{\gamma x^{\alpha}}}.
\end{equation}
The average rank can be computed via derivative of the generating function,
\begin{equation}
    \langle c(x,t) \rangle = \partial_z G(z,x,t)|_{z=1} = \frac{\gamma +\left(2 \gamma  \left(x^{\alpha }-1\right)+1\right) e^{\gamma  t \left(2 x^{\alpha }-1\right)}-1}{\gamma  \left(2
   x^{\alpha }-1\right)} 
\end{equation}

\subsection*{Product Kernel}

In terms of the generating function $G(z,x,t) = \sum_c z^c n(c,t)$, upon multiplying by $z^c$ and summing over $c$ both sides of Eq. \eqref{eq:general_elo_chess} for $\beta = 1$, for $K(c,c') = cc' $ , $v(c)=1$, the dynamics of $G(z,t)$ is,

\begin{equation}
\label{eq:generating_elo}
\partial_t G(z,x,t) = \left[x^{\alpha}z^{2} + (1-x^{\alpha}) - z\right]\partial_z G(z,x,t) \int_0^1 \text{d}y \, P(y)\left[\partial_zG(z,y,t) + G(z,y,t)\right]_{z=1} \, ,
\end{equation} 

Eq. \eqref{eq:generating_elo} is a partial differential equation for the generating function, which depends on the zeroth and first order moments. We rescale time as $\tau =  \int_0^{t} dt' \int_0^{1} dy P(y)  \left[\partial_zG(z,y,t) + G(z,y,t)\right]_{z=1}$. As outlined in the main text we take a uniform distribution of attractiveness.   Eq. \eqref{eq:generating_elo} in the rescaled time $\tau$ is,

\begin{equation}
\partial_\tau G(z,x,\tau) = \left[x^{\alpha}z^{2} + (1-x^{\alpha}) - z\right]\partial_z G(z,x,\tau) \, .
\end{equation}
This equation is the partial differential equation for the generating function of a linear birth-death processes. The generating function is given as previously by the method of characteristics,

\begin{equation}
\label{eq:generating_gel}
  G(z,x,\tau) = \left[\frac{(1-x^{\alpha})(1-z) -e^{-(2x^{\alpha}-1)  \tau} (1-x^{\alpha}(1+z))}{x^{\alpha}(1-z) -e^{-(2x^{\alpha}-1) \tau} (1-x^{\alpha}(1+z))}\right]^{c_0}
\end{equation}

from which the density of users in the condensates is $n(c=0,x,t) = G(z=0,x,\tau)$,
\begin{equation}
\label{eq:generating_gel_zero}
n(c=0,x,\tau) = \left[ \frac{(1-x^{\alpha}) - (1-x^{\alpha})e^{\tau(1-2x^{\alpha})}}{ x^{\alpha} -(1-x^{\alpha})e^{\tau (1-2x^{\alpha})}}\right]^{c_0}
\end{equation}

In order to derive the moment generating function in real time, we need to couple it with the equations,

\begin{equation}
\begin{split}
    & \partial_t  \left[\partial_zG(z,x,\tau)\right]_{z=1} = (2x^{\alpha}-1)\left[\partial_zG(z,x,\tau)\right]_{z=1} \\
     &  \left[ \partial_t  G(z,x,\tau) \right]_{z=1} = 0 \, 
\end{split}
\end{equation}
with solution is $\left[\partial_zG(z,x,\tau) +G(z,x,\tau) \right]_{z=1} = c_0e^{\tau(2x^{\alpha}-1)}+1$. We can then expressed the real and rescaled time as,

\begin{equation}
t(\tau) =  \int_0^{\tau} d\tau' \left( \int_0^1 dy \left[\partial_zG(z,y,\tau') + G(z,y,\tau')\right]_{z=1} \right)^{-1}
\end{equation}
It can be checked that the integral saturates to a constant value for $\alpha>1$. Even though the integral cannot be solve exactly, an approximate asymptotic solution is found as,

\begin{equation}
    \tau \approx \log(\frac{1}{(1+c_0f_{\alpha})e^{-t} -c_0f_{\alpha}})
\end{equation}

with $f_{\alpha}$ a parameters that depends on $\alpha$ and decreases with increasing value of $\alpha$.

Gelation typically occurs with divergence of higher order  moments of users' density $n(c,x,t)$. Moments of order $m$ can be computed from the generating function as,
\begin{equation}
\label{eq:moments_gel}
    \langle c^m (x,t) \rangle = \partial^m_z G(z,t)|_{z=1} \, ,
\end{equation}
In \textbf{Fig S1}  we plot the first three moments, which, as expected, diverge at finite times for $x>x_C$.

\setcounter{figure}{0}
\makeatletter 
\renewcommand{\thefigure}{S\@arabic\c@figure}
\makeatother

\begin{figure}[ht]
    \centering
    \includegraphics[width = 0.5\textwidth]{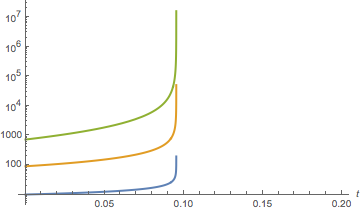}
    \caption{Divergence of the first (blue), second (orange), third (green) moment of Eq.\eqref{eq:moments_gel}. We set $x=0.8,\alpha = 2, c_0 =10$. We set $f_\alpha =1$ as the qualitative behaviour is unchanged by the precise value.}
\end{figure}

 %TC:endignore
 
 %\end{linenumbers}

\end{document}